# Enabling Multiple QR Codes in Close Proximity

**Mercan Topkara**[1]   **Thomas Erickson**[2]   **Umut Topkara**[1]   **Chandrasekhar Narayanaswami**[2]

[1]JW Player, New York, NY

[2]IBM T. J. Watson Research Center, Yorktown Heights, NY

mercan@jwplayer.com, snowfall@us.ibm.com, umut@jwplayer.com, chandras@us.ibm.com

**ABSTRACT**

Quick response codes – 2D patterns that can be scanned to access online resources – are being used in a variety of industrial and consumer applications. However, it is problematic to use multiple QR codes in close proximity: scans can fail or result in access to the wrong resource. While this problem is, strictly speaking, due to the design of the scanning software, the very large number of extant scanning applications makes changing the software a difficult logistical challenge. Instead, we describe the design of a new type of QR code that not only enables the use of multiple QR codes in close proximity, but also is compatible with existing scanning solutions. In an evaluation with 20 users, it was found that the new QR codes were as usable as traditional ones, and that they were superior for selecting one code from many. Users did have initial difficulty in discovering how to use the new QR code, so further work is required on that front. We conclude with a discussion of the pros and cons of pQR codes.

**ACM Classification Keywords**
H.5.2. Information interfaces and presentation (e.g., HCI): User interfaces -- Input devices and strategies.

**INTRODUCTION**

A quick response code (QR code®*) is a 2-D symbol that enables users to connect to an online resource by using an optical scanner to recognize the code (Figure 1). Originally developed for use in the automotive industry in the early 1990's, QR codes have been standardized and are available for general use without license. With the increasing adoption of camera-equipped smartphones capable of scanning images, QR codes have seen increasing use in the consumer space, appearing on products, signage and documents, as well as undergoing more fanciful uses on coins, gravestones, pizzas, and as tattoos. Furthermore, ordinary users can generate their own personal QR codes to enable access to web sites, create calendar events and initiate a call, email, SMS, etc.

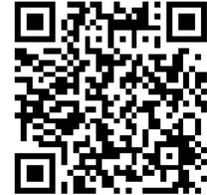

**Figure 1. A QR code**

*This report was released in October of 2015. It describes work done at the IBM T. J. Watson Research Center in 2012 and 2013.*

The future of QR codes in the consumer space is not assured. While advocates point to a growing active user base of nearly 6 million [14], this is a small fraction of smart phone users. At the same time, detractors have suggested that QR codes are dead, or at least not catching on at an impressive rate (e.g. [15, 16]). While a definitive conclusion seems premature, it is worth examining the problems noted. First, marketers have been ridiculed for placing QR codes in difficult-to-scan places – billboards, moving vehicles, and too-quick-to-scan segments of videos – and having QR codes take their smartphone-using users to sites that are not mobile-friendly. A second issue is that barcode scanners are not a default app on most smart phones, thus potential users have to find and download an app. Finally, once on the phone, the scanning app has to be launched, a small but significant effort that can deter casual use. In our view the misuse of QR codes by marketers will correct itself as they gain experience with the medium; and while the need to download and launch QR code scanners is a real barrier, it is not one that has prevented many other mobile apps from becoming quite successful. Even should these limitations prove fatal to the use of QR codes in the consumer space, note that they do not apply to QR code use in industrial settings, where QR codes are alive and well.

In this paper, we take on a different shortcoming of QR codes, one that holds for both consumer and industry applications. As we shall describe in more detail, difficulties can arise when multiple QR codes are used in close proximity to one another, or when they are used near UPC codes. Because of the nature of QR codes, and the systems used to scan them, using multiple scan codes near one another can result in mis-scans or in spurious scans. While this problem could be addressed by making changes to the scanning software, a difficulty is that there are a very large number of scanning systems in existence – into the thousands if one includes smartphone apps in the consumer space – that would need to be changed. Instead of changing the scanning software, this paper describes the rationale, design and evaluation of a new type of QR code. The new code is compatible with existing scanning solutions but enables the use of multiple QR codes in close proximity, opening the door to new ways of using QR codes.

This paper begins by discussing prior work. Next we describe the problem that initiated our work, and the resulting design. In the third part of the paper we evaluate the design. The evaluation addresses two issues: it compares the usability of the new QR code with conventional QR codes, using real world test materials (with the new codes substituted for conventional codes, as appropriate); and it examines various approaches to instructing users on how to use the new codes. In addition, the evaluation provides insights on how people physically manipulate their phones when scanning QR codes, and some of the problems they encounter. After the evaluation we discuss our findings, looking at how well the new type of code fared in terms of usability, learnability and discoverabiliy. Finally, we discuss the pros and cons of pQR codes, and some implications of this work for the future. While our work was driven by very near-term issues, it also has implications – and raises interesting questions – for those exploring the domain of wearable displays and embodied interaction.

**BACKGROUND**

QR codes were first introduced in auto industry by DENSO [6]. Compared to existing 2D codes, the QR code was designed to carry more data, and to be robust to damage, with error correction ranging from 7% – 35%. They can also be scanned from different angles, and newer versions can compensate for distortion due to their being displayed on curved surfaces. Although DENSO patented the QR code, it declared it would not exercise its patent rights, and encouraged free use and standardization of the code. This has resulted in widespread adoption of the QR code in industry and consumer applications, although as noted in the introduction there is debate about the degree to which QR codes will prosper in the consumer space.

There is a large body of research that examines the learnability and usability of various mobile phone-based interaction techniques. For instance, Ballagas et al. [3] surveys 10 mobile phone-based positioning techniques and provide a comparison of these techniques on several ergonomic measures such as cognitive load, motor load, fatigue, error proneness etc. Similarly Ruzkio, et al. [13] empirically evaluated a range of techniques and offered a summary of the advantages and disadvantages of touching, pointing, scanning and user-mediated interactions.

With respect to QR code use, Ballagas et al. [2] state that this type of interaction has medium cognitive load, high motor load and high error-proneness compared to low cognitive load and error proneness of controlling GUI widgets (marked with 2D codes) on a digital display using camera phones, which is also in line with Ruzkio, et al.'s comments on scanning [13]. Similarly, Toye et al [17] describe a usability and discoverability study on 2D codes (similar to QR codes) scanned by camera phones. In this study, users performed several point and scan tasks that measured how quickly they could discover how to use them in a simulated real-world application (checking into lines for rides in a theme park). They also studied how quickly and accurately novice users were able to click on visual tags of varying sizes. Their results were positive: they observed that all users figured out how to navigate in the 'theme park' using codes within 15 minutes, and that their accuracy and response time were within acceptable ranges in the point and scan tests.

Besides their increasing use in various consumer-oriented applications, QR codes have been the subject of considerable work in HCI and mobile computing. User groups targeted have ranged from children [4] to the cognitively impaired [5] to the chronically homeless [1]. QR codes have also been proposed as means for the sharing of content on public displays [8], supporting adaptive learning [12], extending augmented reality systems [7] and delivering government services [9].

These new applications have used QR codes as they are – they have not focused on modifying the design of QR codes. QR codes have been relatively stable in terms of their makeup. The standardization process has included the development of different versions of QR codes that support different levels of error correction and increasing robustness. Beyond that, the only alterations to QR codes have been the development of micro-QR codes, encrypted QR codes, and QR codes that enable the use of colors and embedded pictures to support the needs of those concerned with branding [6].

**DESIGN**

**The Problem…**

Our initial encounter with the proximity problem occurred when we were working with a client who wished to augment a marketing campaign by placing a QR code on their product's packaging, alongside the UPC barcode. To their dismay, they discovered that the cash register scanners used to scan the price from the UPC barcode could also scan the QR codes – and that when they detected two codes they would randomly choose one. This type of error – which was quite common when the UPC and QR codes were printed near one another – resulted in either an explicit scan error, the silent addition of zero to the running total, or the addition of the wrong product to the running total. While it would be possible to redesign the product packaging to put the QR code on one face of the package, and the UPC code on the other, this was a significant design issue that impacted the branding of the product.

Further investigation showed that similar problems occurred when two or more QR codes were used in proximity to one another. Most QR code scanning software assumes that a single code is present, and will automatically search for and acquire a code. If multiple QR codes are present, they can easily fit within one scanning screen (Figure 2): in that case, the scanning software will make random choice (or at least a choice the user has no control over).

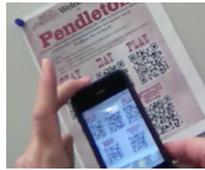

**Figure 2. Multiple codes in proximity.**

While most QR codes are used singly, it is not difficult to find examples of the use of multiple codes in proximity (Figure 3). While this is more testimony to the lack of literacy with the medium that we've already remarked upon, it also shows a desire to juxtapose multiple codes.

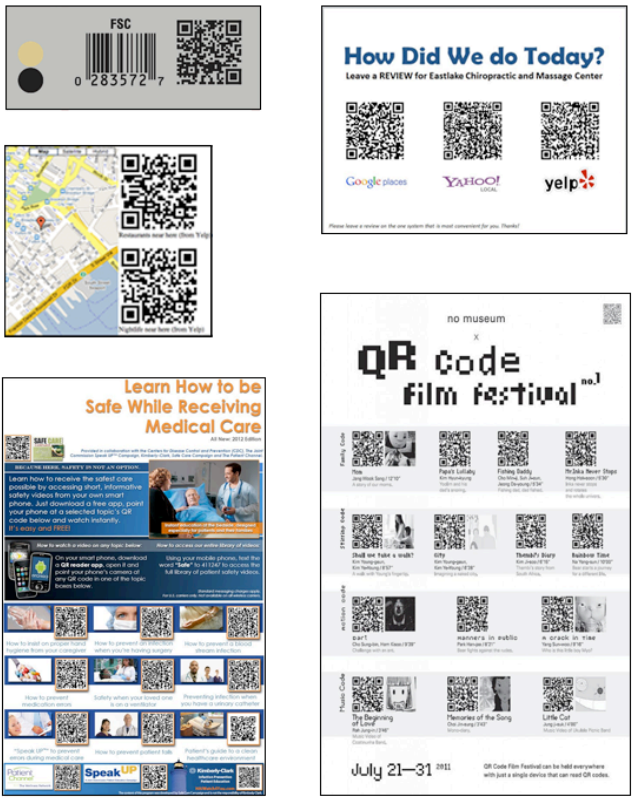

**Figure 3. Examples of juxtaposed scan and QR codes.**

It would be a simple matter to address this problem at the software level –when multiple codes are detected ask the user to select one – were it not for the great diversity of scanning systems and scanning applications. No overall count is available, but there are easily hundreds of scanning applications in the smartphone space alone. Similarly, in the retail (checkout register) and industrial (inventory tracking) domains there are many scanning systems as well, and the system does not always have affordances for its users to indicate a selection. So, instead of addressing the problem at the software level, we developed a solution that was software independent.

**…and a Solution**

The solution we arrived at was to 'damage' the QR code in a way that enabled it to be easily repaired via human intervention. Specifically, recognition of a QR code can be defeated by inserting particular patterns in one corner of the code, and then the damage can be undone by having the user cover the damaged corner of the code with a finger, or another opaque object.

In a little more detail, scanning a QR code begins with the detection of three "Finder Patterns," one on each of three corners of the QR code. The Finder Patterns are the large black-white-black squares, and help identify the presence and orientation of a QR code in an image; alignment patterns, smaller black-white-black squares, assist in straightening out QR codes placed on curved surfaces (Figure 4a).

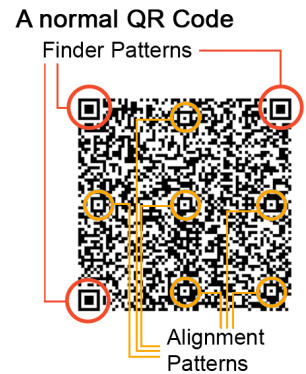

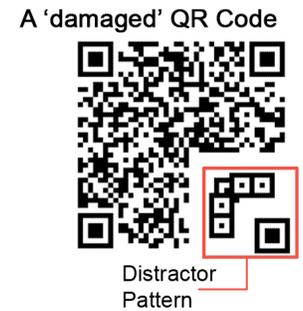

We damage the QR code by adding a "Distracter Pattern" – a fourth Finder Pattern and a cluster of Alignment Patterns – to the corner of the QR code without a Finder Pattern (Figure 4b). This has the effect of greatly enlarging the search space, and rendering the code unscannable, until the user covers the corner with a finger. While the damage does obscure some information present in the QR code, the information redundancy in QR codes is sufficient that the damage is not irreversible: the size of the area covered by the distracter pattern can be adjusted to be within the error tolerance of the version of the QR code being produced.

**Figure 4. (a) Finder and Alignment Patterns in a normal QR code; (b) a QR code 'damaged' with a distractor pattern.**

We refer to this new type of QR code as a "peacocked" QR (pQR) code, by analogy to the peacock's use of aposematism. Aposematism is a signal that some animals use to advertise their undesirability as a prey, usually with bright colors, patterns, sounds and odors. The eyespots on the peacock's tail are thought to be an aposematic adaptation, since animals have a perceptual bias towards recognizing eyes. Since we exploit the QR reader's mechanism for recognition of the Finder Patterns, referring to the new codes as "peacocked" seemed apt and memorable.

*Designing Instructions*

Assuming that this approach actually works to produce a code that is unscannable until a human intervenes – an issue we will address in the evaluation portion of the paper – we are left with one question: how do we make it evident to users that they need to intervene, and how do we show them how to intervene? In the early stages of design we developed some sketch-based prototypes, that exploited combinations of three features: (i) a curved boundary to signify the to-be-covered area; (ii) an indication that a finger could be used to do the covering, and sometimes (iii) add a brief textual instruction. We decided to use the evaluation to refine the instructions – our hope was to develop visual instructions that were clear enough not to require text. Figure 5 shows the four versions we evaluated.

**THE STUDY**

The study had two aims. The primary aim was to evaluate how well the new 'peacocked QR codes' worked compared to traditional QR codes. Here we were interested in understanding how the two types of codes compared to one another in terms of usability, and in comparing the strengths and weaknesses of the new codes to standard QR codes. A secondary aim was to get a sense of how well our instructional diagrams worked, and to guide us in refining them if necessary.

Overall, the study consisted of four parts: (i) an initial survey that gathered information about the user and his or her prior experience with QR codes; (ii) an instructional phase where the user was presented with a peacocked QR code and an instructional diagram, and asked to discover how to scan the code; (iii) a task phase where the user was asked to scan both traditional and peacocked QR codes using real world examples of multiple-code posters; and (iv) a wrap up phase with a survey about the task phase, followed by debriefing and open-ended discussion.

**Method**

We recruited 20 participants from among colleagues in our research organization via emailed invitations asking if they would volunteer to participate in a study to evaluate the usability of a new type of QR code. The study took place in a meeting room, with our test materials posted on the walls; we provided participants with the same phone and scanning app for use in the evaluation.

We recorded the interaction of participants, with their permission, via using different channels. During the instructional and task phases, participants wore a headset camera attached to their right ear so that we could see what they were looking at. We also recorded via a fixed tabletop camera – this device recorded audio of the whole process. Finally, we captured a screen recording of the screen of the phone used to do the scanning.

*Phase 1 – Gathering background information*

Participants were welcomed, reminded of the purpose of the study, and asked to fill in a background survey. The survey enquired about their handedness, and their experience in using smartphones in general and QR codes in particular.

*Phase 2 – Assessing the discoverability of the pQR code*

As already discussed, we knew that providing instructions on how to use the pQR codes would be an issue, and we decided to use this phase of the evaluation to try variations on the instructional diagrams. We told each participant that this was a new type of QR code, and that they should try to figure out how to scan it without our help. We gave participants a sheet of paper containing the to-be-scanned code, and a small instructional diagram. If the participant was not able to figure out how to scan the pQR code within 5 minutes, we provided an increasingly explicit series of hints until they were able to successfully scan the code.

We had initially intended to test two diagrams, but as participants encountered considerable difficulty with the first two versions of the instructions, we designed more literal diagrams with captions. We will discuss the results of this Phase in detail in the following sections, however we would like to note here that most users immediately understood the instruction to cover the corner of the code, however difficulties arose in application as many users' intuition was to touch and cover the corner on the image of the code on the phone screen instead of occluding the corner of the pQR code printed on the paper or look for more cues within scanner app. The resulting set of four instructional diagrams is shown in Figure 5.

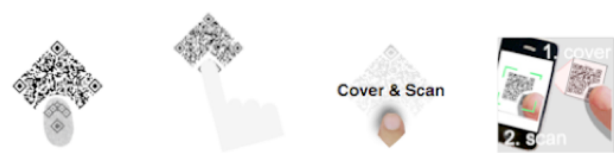

**Figure 5. Four versions of the instructional diagram**

*Phase 3 – Scanning both QR and pQR codes*

In this phase of the evaluation, participants were asked to scan two versions (QR and pQR) of each of two posters; order of presentation of traditional and pQR codes was counterbalanced across participants. To do the task, participants had to walk around the lab and carry the phone with them; they began each scanning task with the phone sitting on the table in front of them. For the first poster (Figure 6a), we had them scan one code; for the more complex second poster (Figure 6b), we asked them to scan three different codes.

Constructing test cases for comparing systems is a difficult matter, because of the concern that any given test case might be biased towards the strengths of one of the to-be-compared system. For this reason, we worked with already existing examples that used multiple QR codes (Figure 6). For the experiment, we used the original posters with QR codes without any modification, and for each example we created a version of the poster with pQR codes. We also modified the wireless network so the original URLs encoded in the posters took the users to web pages we created, which clearly indicated the scanned option (e.g., 'eat'). The only other variation that was made is that we present pQR codes in a 'diamond' orientation, so that the to-be-covered corner appears at the bottom of the code, and does not favor right or left handers. After the above substitutions, the posters were reproduced at their original size, and posted on the wall in the room where the evaluation took place.

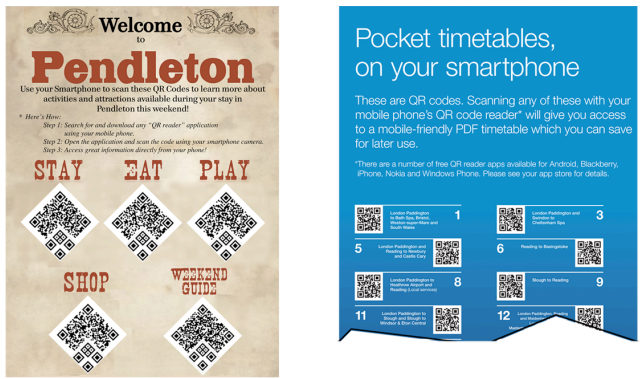

**Figure 6. Multi-code test posters:**
**(a) Pendleton guide (pQR version; QR version not shown);**
**(b) Transit schedule (QR version; pQR version not shown)**

*Phase 4*
After the task phase, participants filled in a survey that assessed their experience during this phase. Then responded to Likert scale questions about the usability of QR and pQR codes, and their learnability in general and for the multiple scan tasks. The participants were then debriefed, and engaged in general discussion about their impressions of the two types of codes, and how they might be improved.

## RESULTS
Twenty people completed the experiment. According to the initial survey, all but two had owned a smart phone for two years or longer, and all but one were right handed. With respect to QR code use, 35% had never scanned a QR code (though some reported clicking on them on the web); of the rest, about half reported using them a few times a year or more – only two of those reported weekly or daily use of QR codes. 40% were female and 60% male; 65% were in the 26-35 age range.

**Usability**
In general, pQR codes were comparable to QR codes in their usability. After the task phase of the evaluation, participants answered survey questions on the general usability of QR and pQR codes – their responses are shown in Table 1, grouped into agree (strongly agree or agree), neutral, or disagree (strongly disagree or disagree) categories. Response patterns for QR and pQR codes were similar, with majorities feeling they are not confusing and that they are easy to use and intuitive. The only distinct difference is that users appeared less concerned about making errors with pQR codes than traditional QR codes (75% vs. 45% *disagreed* the corresponding codes were prone to errors).

Because many of our participants had little or no experience with traditional QR codes according to the initial survey, we also looked just at the responses of the 10 users who reported interacting with QR codes a few times a year (or more). We found the same pattern of responses shown in Table 1 for them, except that the difference in concern about making errors vanished (70% *disagreed* regular QR codes were error prone, and 80% *disagreed* pQR codes were error prone). Overall, we conclude that QR and pQR codes have about the same level of usability and user-friendliness.

Another survey question asked specifically about the problem we set out to address: scanning codes when multiple codes are in close proximity. Table 2 shows participants' ratings of the overall ease of scanning a code in the multiple code condition: 100% felt that the pQR codes made it easy to select one code from many, versus 60% who felt the same way about traditional QR codes. We will return to this issue in the discussion.

| QR codes are | …confusing | …easy to use | …intuitive | …error prone |
|---|---|---|---|---|
| *Disagree* | **14 (70%)** | 3 (15%) | 3 (15%) | 5 (25%) |
| *Neutral* | 2 (10%) | 4 (20%) | 6 (30%) | 6 (30%) |
| *Agree* | 4 (20%) | **13 (65%)** | **11 (55%)** | **9 (45%)** |
| pQR codes are | …confusing | …easy to use | …intuitive | …error prone |
| *Disagree* | **12 (60%)** | 2 (10%) | 6 (30%) | **15 (75%)** |
| *Neutral* | 4 (20%) | 4 (20%) | 2 (10%) | 4 (20%) |
| *Agree* | 4 (20%) | **14 (70%)** | **12 (60%)** | 1 ( 5%) |

**Table 1. General usability of QR vs pQR codes**

| It was easy to select one code from many… | …using the traditional QR codes | …using one of the new pQR codes |
|---|---|---|
| *Disagree* | **12 (60%)** | 0 ( 0%) |
| *Neutral* | 3 (15%) | 0 ( 0%) |
| *Agree* | 5 (25%) | **20 (100%)** |

**Table 2. Ease of use for selecting one code from many**

To look more closely at the usability of the QR vs. pQR codes, we watched the video to count the number of mistakes participants made when scanning the poster (Table 3). Note that some of the mistakes were due to the participant intentionally scanning the wrong code either because they mis-heard the instructions or because (in the case of the transit poster) the poster design led to some confusion. All mistakes on the pQR posters were due to this type of mistakes. (One instance could not be counted due to video failure.)

For the first poster, participants collectively made a total of 14 mistakes on the QR code version, and 4 mistakes on the pQR version. Or to look at it another way, 9 of the participants made fewer mistakes when scanning the pQR version of the poster. In the case of the second poster, where participants were asked to scan three codes, they collectively made a total of 24 mistakes, compared to 5 mistakes for the pQR version of the poster. For this poster, 8 of the participants made fewer mistakes on the pQR version, although 2 made few mistakes on the QR version. These findings are consistent with participants' ratings of the ease of selection shown in Table 2.

|  | **Mistakes for QR poster version** | **Mistakes for pQR poster version** |
|---|---|---|
| *Pendleton* | **14** | 4 |
| *Transit* | **24** | 5 |

**Table 3. Collective mis-scans of QR vs. pQR poster versions**

In summary, the pQR codes appear to work well in terms of usability. Overall, they are comparable to regular QR codes in their ease of use, and they appear to be superior with respect to the situation for which they were designed: selecting one code from among others in proximity to it.

**Ease of Learning**
Now let us turn to the question of how easy it is to learn to use the new pQR codes. There are two issues here: learnability and discoverability. Learnability is the most general issue, and it has to do with how difficult it is to develop the skills to actually make use of the pQR code. Learnability matters regardless of whether the code is being used in an industrial setting or in a public consumer setting. Discoverability has to do primarily with the consumer space: that is, when someone encounters a pQR code for the first time, how difficult is it to figure out what to do?

*Learnability*
On the survey, 75% of our participants agreed that the pQR code was easy to learn; 15% were neutral and 10% disagreed. The survey also asked participants how easy it was to learn to select one code from many, for both the QR and pQR codes. Results are shown in Table 4: 95% of the participants agreed that the pQR code was easy to learn; in contrast, 55% agreed it was easy to learn to do the same with regular QR codes.

| It was easy to learn to select one of the | …traditional QR codes from the others | …new pQR codes from the others |
|---|---|---|
| *Disagree* | 3 (15%) | 0 ( 0%) |
| *Neutral* | 6 (30%) | 1 ( 0%) |
| *Agree* | **11 (55%)** | **19 (95%)** |

**Table 4. Ease of learning for selecting one code from many**

While pQR codes were clearly preferred for learnability, we had not expected traditional QR codes to perform so well. However, as we observed users doing the experimental task, we saw that many who had initial problems scanning one regular QR code in close proximity to another developed workarounds after a couple of failures. One user developed the technique of pointing the phone at the floor, bringing it close to the code he wished to scan (still pointing at the floor) and then rapidly rotating it upward 90 degrees when it was close enough that the code filled the scanning field of the camera. Other users would cover the camera of the phone with their hand, position it in front of the code they wanted to scan, and then remove their hand; however, with this technique, their hand and phone tended to occlude the code they were trying to locate. See Figure 7 for a screenshot of recording of one user using this technique to avoid an unintentional scan. The bottom line is that while most users initially found scanning ordinary codes in the close proximity condition cumbersome, some developed effective solutions.

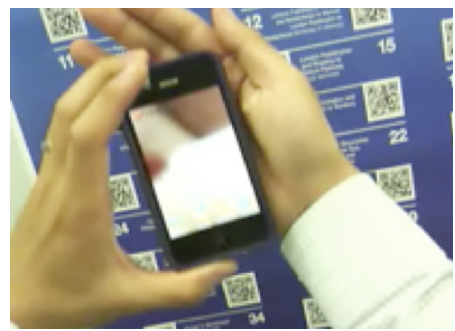

**Figure 7 One of the techniques developed by users to avoid unintentional scans when QR codes are located in close proximity**

*Discoverability*

The other ease of learning issue was discoverability. This was addressed in phase 2 of the study, when we told participants that they needed to figure out how to scan a new type of pQR code, and gave them a piece of paper with the code and one of four instructional diagrams. This proved more difficult than we had anticipated, with the majority of participants taking several minutes to discover how to scan the pQR codes, and some eventually needing to be shown.

Three types of problems occurred. First, some participants simply ignored the instructions, and started experimenting with the smart phone and scanner app. Second, some participants (often after experimenting for a bit) looked for help in the smartphone scanner application – however that was a generic scanning application, and had no information about how to use QR codes. Third, and most interesting, a number of participants viewed the instructions, realized that they needed to cover the corner of the pQR code with their finger, but focused on covering the corner on the image of the pQR code shown on the screen of the phone, rather than the actual pQR code that they were scanning. See Figure 8 for screenshots of a recording of two different users covering the corner on the phone screen.

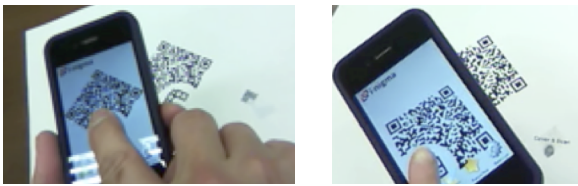

**Figure 8 Several users' first intuition was to cover the 'distracter' corner on the phone screen**

It *is* the case that the third and fourth versions of the instructional diagrams – which included both text and more explicit depictions of fingers – were better in terms of time to doing a successful scan, and fewer people needing hints. However, as we only had 5 people for each version of the instructions, we cannot confidently say which version is superior. It is clear that this is an area where further work is needed.

## DISCUSSION AND CONCLUSION

In this paper we described a problem and proposed a solution. The problem is that QR codes, when used in close proximity to one another, are subject to scanning errors due to the design of scanning applications. The scanning applications automatically scan and recognize QR codes, and if multiple codes are present will pick one, seemingly at random. While there are a number of obvious solutions, such as building a selection mechanism into the scanning application, the difficulty is that there are thousands of scanning applications in existence.

The solution described in this paper is software independent. It involves rendering the QR code unreadable by inserting additional "finder patterns" in one corner; we refer to this as a "peacocked" QR code, alluding to the multiple eyespots on a peacocks tail. The pQR code can be read by the simple expedient of the user placing his or her finger over the 'peacocked' corner of the pQR code – at that point, it can be read by any QR code reader. Although inserting the distracter pattern in the corner of the QR code obscures some of its information, QR codes are designed with enough redundancy that they still scan correctly.

The pQR code was evaluated in a study with 20 participants using real examples of posters that used multiple QR codes. Participants rated both QR and pQR codes as quite usable and learnable; however, while all agreed that pQR codes made it easy to select one code from many, only 65% agreed QR codes were easy to use in this case. Observations of error rates in the two evaluation tasks were consistent with these ratings. Thus, we have a solution to the multiple QR scanning problem that can be implemented independently of software, simply by changing the visual design of QR codes.

As participants pointed out in discussions towards the end of the evaluation, pQR codes have some limitations. The most common comment was that pQR codes require the use of two hands – one to hold the phone, and one to point to (and cover the corner of) the pQR code – and that can be cumbersome or inconvenient in some circumstances. Participants also noted that it might be difficult to point to pQR codes: the code might be inconveniently located, there might be a crowd around the poster or sign that could make it difficult to get close enough, or the code might be far away (although as a couple of participants noted, one could still occlude the corner of the code by positioning one's finger in front of the camera lens). Finally, a few participants remarked that if the pQR code was large, its corner might be too large to cover up.

Those objections aside, nearly all participants agreed that the pQR code solved the multi-code problem quite effectively. While many devised ways of effectively selecting one from many QR codes, most agreed that this was cumbersome. As one said, "*When I was trying to focus* [the scanner] *on the QR code* [on the transit schedule], *I couldn't see the number*" (because his phone was occluding it). In contrast, he went on to say, "*with the pQR code all I had to look for was my finger* [over the pQR code] – *everything else was disabled. There's almost a failsafe there.*" A number of other participants echoed this theme, appreciating the greater sense of control, and of engagement. A few even thought it was fun – like putting a 'button' out in the world.

Perhaps the biggest drawback of the pQR code, at least for use in consumer applications, is that users are accustomed to interacting directly with the software on the phone instead of items from physical world and this biases the discoverability of pQR codes. Even with instructional diagrams, some users took minutes to figure out how to use the pQR code, although once they knew what to do they agreed it was easy. It may be that further work on developing instructional diagrams will eliminate this problem. It is also the case that when used in a public setting, pQR codes are in a sense self-documenting: bystanders will be able to see a pQR code user holding her phone with one hand, and pointing to select the pQR code she wants with the other.

Finally, the gradual shifting of paradigms may also ameliorate the problem. We observed – in the initial discoverability test – that participants were often fixated on interacting only with their phones. As one person said in the final interview, "*I assumed the interaction would be on application, I assumed it would just get camera feedback from physical world, nothing else.*" While this is a reasonable assumption today, it seems likely that this will change. Application paradigms like augmented reality, the advent of wearable displays such as Google Glass that do not provide an overlay touch-input surface, and other systems that feature in-the-world interaction (e.g. [18]), will weaken the assumption that all interaction takes place on the surface of a device.

## ACKNOWLEDGMENTS

We thank our users for participating in our study, and for their willingness to share their time and experiences.

## NOTES
\* QR code® is registered trademark of DENSO WAVE INCORPORATED in JAPAN and other countries.

## REFERENCES
1. Aljadann, A., Bihani, K. and Gebrekristos, M. QR-Codes for the Chronically Homeless. *Ext. Abstracts CHI 2008*, ACM Press (2008), 3879-3883.
2. Ballagas, R., Rohs, M., & Sheridan, J. G. (2005, April). Sweep and point and shoot: phonecam-based interactions for large public displays. *Ext. Abstracts CHI 2005*. ACM Press, 1200-1203.
3. Ballagas, R., Borchers, J., Rohs, M., & Sheridan, J. G. (2006). The smart phone: a ubiquitous input device. *Pervasive Computing, 5*, 1 (2006), 70-77.
4. Ceipidor, U.B., Mdeaglia, C.M, Perrone, A. Marsico, M. and Romano, G. A museum mobile game for children using QR-Codes. *Proc. IDC, 2009*. ACM Press (2009), 282-283.
5. Chang, Y.J., Tsai, S.K., Chang, Y.S. and Wang, T.Y. A Novel Wayfinding System Based on Geo-coded QR codes for Individuals with Cognitive Impairments. *Proc. ASSETTS '07*. ACM Press (2007), 231-232,
6. Denso ADC. QR code Essentials. Denso ADC (2011). http://www.nacs.org/LinkClick.aspx?fileticket=D1FpVAvvJuo%3D&tabid=1426&mid=4802 Accessed 15-Sept-2013.
7. Fuhrt, B. (Ed). *Handbook of Augmented Reality*. Springer, 2011.
8. Geel, M., Huguenin, D. and Norrie, M.C. PresiShare: Opportunistic Sharing and Presentation of Content Using Public Displays and QR codes. *Proc. PerDis '13*. ACM Press (2013), 103-108.
9. Lorenzi, D., Vaidya, J., Chun, S., Shafiq, B., Nabi, G. and Atluri, V. Using QR codes for Enhancing the Scope of Digital Government Services. *Proc. dg.o 2012*. ACM Press (2013), 21-29.
10. Mäkelä, K., Belt, S., Greenblatt, D., Häkkilä, J. Mobile interaction with visual and RFID tags: A field study on user perceptions. *Proc. CHI 2007*. ACM Press (2007), 991-99.
11. Rohs, M. Real-world interaction with camera phones. *Ubiquitous Computing Systems*. Springer (2005), 74-89.
12. Rouillard, J. and Laroussi, M. PerZoovasive: Contextual pervasive QR codes as tool to provide an adaptive learning support. *Proc. CSTST '08*. ACM Press (2008), 542-548.
13. Rukzio, E., Broll, G.B., Leichtenstern, K. and Schmidt, A. Mobile Interaction with the Real World: An Evaluation and Comparison of Physical Mobile Interaction Techniques. *Proc. AmI '07*. Springer-Verlag (2007), 1-18.
14. Scanlife. Trend Report 2013 Q2. http://www.scanlife.com/trend-reports. Accessed August 29, 2013.
15. Stout, A. The death of the QR code. Marketing Land, April 4, 2013. http://marketingland.com/the-death-of-the-qr-code-37902. Accessed August 28, 2013.
16. Taub, A. QR codes Are Dead! Long Live QR codes! A Conversation With Scan's Founder, Garrett Gee. *Forbes*, Dec 6, 2012. Accessed August 29, 2013.
17. Toye, E., Sharp, R., Madhavapeddy, A., Scott, D., Upton, E., and Blackwell, A. Interacting with mobile services: An evaluation of camera-phones and visual tags. *Personal and Ubiquitous Computing, 11*, 2, (2007), 97-106.
18. Zhou, Y., Xu, T., David, B., & Chalon, R. Innovative wearable interfaces: an exploratory analysis of paper-based interfaces with camera-glasses device unit. *Personal and Ubiquitous Computing*, (2013), 1-15.